\documentclass[aps,amssymb,amsfonts,preprint,floatfix,showpacs,superscriptaddress,10pt]{revtex4-1}

\usepackage[utf8]{inputenc}
\usepackage{bm}
\usepackage{xcolor}
\usepackage{siunitx} 
\usepackage[colorlinks=false, pdfborder={0 0 0}]{hyperref}
\usepackage{cleveref}
\usepackage{todonotes} %https://www.ctan.org/pkg/todonotes
\usepackage{abraces}% http://ctan.org/pkg/abraces % braces over the text

\crefname{figure}{Figure}{Figures}
\crefname{equation}{Equation}{Equations}
\crefname{table}{Table}{Tables}

\newcommand{\gr}[1]{\textcolor{black}{#1}}
\newcommand{\jk}[1]{\textcolor{black}{#1}}

\newcommand{\sg}[1]{\textcolor{black}{#1}}

\DeclareSIUnit\pxl{pixels}
\DeclareSIUnit\ppm{ppm}
\DeclareSIUnit\ppb{ppb}
\DeclareSIUnit\ppt{ppt}
\DeclareSIUnit{\liter}{$\ell$}

\graphicspath{{./figures/}}

\begin{document}
\title{Shifting velocity of temperature extremes under climate change}

\author{Joan Rey}
\affiliation{Institute for Environmental Sciences, University of Geneva, bd Carl Vogt  66, 1211 Geneva 4, Switzerland} 

\author{Guillaume Rohat}
\affiliation{Institute for Environmental Sciences, University of Geneva, bd Carl Vogt  66, 1211 Geneva 4, Switzerland} 
\affiliation{Faculty of Geo-Information Science and Earth Observation, University of Twente, 7514 AE Enschede, The Netherlands}
\affiliation{National Center for Atmospheric Research (NCAR), 80301 Boulder, CO, United-States} 

\author{Marjorie Perroud}
\affiliation{Institute for Environmental Sciences, University of Geneva, bd Carl Vogt  66, 1211 Geneva 4, Switzerland} 

\author{Stéphane Goyette}
\affiliation{Institute for Environmental Sciences, University of Geneva, bd Carl Vogt  66, 1211 Geneva 4, Switzerland} 
\affiliation{Group of Applied Physics, University of Geneva, GAP, 22 chemin de Pinchat, 1211 Geneva 4, Switzerland}

\author{J\'er\^ome Kasparian}
\email{jerome.kasparian@unige.ch}
\affiliation{Institute for Environmental Sciences, University of Geneva, bd Carl Vogt  66, 1211 Geneva 4, Switzerland} 
\affiliation{Group of Applied Physics, University of Geneva, GAP, 22 chemin de Pinchat, 1211 Geneva 4, Switzerland}

\vspace{10pt}

\date{\today}

\begin{abstract}
Rapid changes in climatic conditions threaten both socioeconomic and ecological systems, as these might not be able to adapt or to migrate at the same pace as that of global warming. In particular, an increase of weather and climate extremes can lead to increased stress on human and natural systems, and a tendency for serious adverse effects. We rely on the EURO-CORDEX simulations and focus on the the screen-level daily mean temperature (T2m). We compare the shifting velocities of the cold and hot extremes with these of the associated central trends, \emph{i.e.}, the arithmetical mean or median.  
We define the extremes relative to the T2m distribution as it evolves with time over the period of 1951--2100. We find that temperature extremes shift at a similar velocity compared to that of the central trends. Accordingly, the T2m probability distribution shifts mostly as a whole, as the tails of the distribution increase together with the central trends. 
Exceptions occur however in specific regions and for the clustering of warm days, which shifts slower than all other extremes investigated in this study. 
\end{abstract}

\maketitle

\section{Introduction}

Global warming is arguably one of the most pressing contemporary societal issues. Changes in both global and local climatic conditions  directly affect a number of natural and anthropogenic systems worldwide~\cite{RN955}. There is a vivid demand from stakeholders, decision makers, and other practitioners for climate change projections in view of both mitigation and adaptation strategies~\cite{RN2716}. 

The adaptation window for both ecological and human systems is highly dependent on the pace of climate change. When the focus is on the local adaptation of ecosystems, industrial process or organisational changes, such a speed is locally defined using the temporal derivative of the considered climatic variables.
However, when adaptation implies migration of species, shifts of biota, or relocation of human settlements and activities, spatial aspects of the climate shift are crucial. Two  approaches have been employed so far to explore the potential impacts of such spatial aspects on various socio-economic activities, such as forestry~\cite{RN2721}, agriculture~\cite{RN2723,RN2722}, and urban planning~\cite{RN2718,RN2724}.

On the one hand, climate analogues (or "climate twins")~\cite{RN2717,RN2725,RN2720} are well-suited to raise awareness of decision makers and lay audience about the speed and magnitude of climate change, as they convey an easy-to-understand message by comparing climatic conditions at well-identified places and times. Climate analogues are areas that are expected to experience, for a given time-period, the climate of a reference location at another time period. 
Dividing the distances between the two paired locations by the considered time interval defines a shifting velocity of climate~\cite{RN2719,RN2720}. 
This approach allows to consider multiple variables, such as temperature and precipitation~\cite{RN2729} or purpose-specific climate indices, $e.g.$ bio-climatic indices~\cite{RN2722,RN2730}. It also accounts for natural barriers such as mountainous regions or marine areas. Furthermore, it helps identifying climates that are expected to disappear in the future, or future emerging conditions not encountered to date -- $i.e.$ those that have no future or past analogues~\cite{RN2726,RN2728}. However, the shifting velocity may depend on the matching scheme and parametrization, and can be overestimated~\cite{RN2788,RN2789}.

Alternatively, one can focus on the shifting velocity of climate through the displacement of isopleths, \emph{e.g.}, isotherms if one focus on temperature. Such shifting velocity can be defined as the displacement vector, per unit of time, of the local climatic conditions characterized by the isopleth(s) corresponding to the considered variable(s)~\cite{Loarie2009}. For a single variable, tracking isopleths motion is equivalent to seeking analogues with zero tolerance. Determining shifting velocities from the local temporal evolution and the  gradient is achieved at the cost of implicit assumptions, as detailed and discussed below. Within this framework, a system can keep pace with a moving climate if its maximum shifting velocity is at least comparable to that of the relevant climatic parameters~\cite{Loarie2009}. 

Among all climate-related phenomena, extreme temperature events~\cite{RN2753} have a wide range of impacts on anthropic and natural systems. They affect both directly and indirectly human health and well-being~\cite{RN2754}.
Examples include increasing heat-related illness and casualties~\cite{RN2733,RN2744,RN2799}, power failures, degradation of critical infrastructure during heat waves~\cite{RN2791}, reduced crop yields~\cite{RN2743,RN2799}, and economic loss -- $e.g.$ due to reduced labour hours, extreme events-related costs, or increased health care needs~\cite{RN2790}. 
The expected impact of climate extremes is made even more important by the projections that some of these extreme events will become more frequent, more widespread, longer, and/or more intense during the $21^\textrm{st}$ Century~\cite{RN955,RN70,RN2735,RN2736,RN2734}.  New insights about the spatial and temporal shift of temperature extremes could  help assess the effect of climate change on the spatial behavior, occurrence, and distribution of these extremes, to better assess the future state of ecosystems ($e.g.$~\cite{RN2745,RN2746,RN2747,RN2748,RN2749}) and to anticipate impacts and required adaptation measures for human populations and systems ($e.g.$~\cite{RN2718,RN2724,RN2717,RN2725,RN2719,RN2750}).

Here we quantify the spatial shift of temperature extremes over Europe and compare their velocity to that of the corresponding central trends. While the climate is defined by numerous variables, temperature change is considered as a reasonable proxy of climate change, and its projection bear much less uncertainties and noise than, e.g., precipitation or wind. 
We focus on screen-level air temperature (T2m) over the European Coordinated Downscaling Experiment (EURO-CORDEX) domain~\cite{RN2216}, for the period 1951--2100. We consider the climate scenario according to the Representative Concentration Pathway (RCP)~8.5~\cite{RN2755}. Under this high greenhouse gas emission scenario, the increase in mean T2m ranges from \SIrange{2.5}{5.5}{\celsius} over continental Europe by 2081--2100 relative to 1986--2005~\cite{RN2808}. 

Up to 27 different definitions of extreme temperature events have been proposed by the World Meteorological Organization~\cite{RN2794,RN2795,RN2796,RN2797}. We focus on several of them: the tails of the probability distribution function (PDF) (high/low quantiles or standard deviations), the exceedance of fixed thresholds (tropical nights or frost days), and the clustering of warm days, in an approach similar to  warm spells~\cite{RN2792}.

We find that, overall, temperature extremes and the corresponding central trends shift across Europe with a similar velocity. This applies to extremes defined relative to the PDF tails as well as to a fixed threshold. In contrast, the clustering of successive warm days shifts much slower than the central trend. We also identify and discuss the few regions where the shifting velocity of the temperature extremes deviates from that of the central trends. 
Finally, we show that the similarity of shifting velocities between extremes and central trends is closely related to the definition of new normals~\cite{RN2732}, i.e., to the fact that the mean temperatures are increasing while the other parameters of the distribution are not significantly changing.

\section{Methods}
\subsection{Datasets}

This work uses high-resolution (\SI{0.11}{\degree}) regional climate models (RCMs) outputs using the RCP 8.5 scenario~\cite{RN2755} within the EURO-CORDEX~\cite{RN2216} initiative over the period 1951--2100. This long period maximizes the climate evolution and therefore the potential discrepancies between shifting velocities of central trends and extremes. The computational domain covers the following geographical coordinates: (\SI{63.55}{\degree}N, \SI{51.56}{\degree}W), (\SI{63.65}{\degree}N, \SI{72.56}{\degree}E), (\SI{20.98}{\degree}N, \SI{14.31}{\degree}W), and (\SI{20.98}{\degree}N, \SI{72.56}{\degree}E). The ability of EURO-CORDEX simulations to reproduce present-day temperature extremes has been demonstrated ($e.g.$~\cite{RN2151,RN2129}) and their outputs have been widely used to analyze projections of extreme temperatures in Europe ($e.g.$~\cite{RN2737}). We use historical runs for the period 1951--2005 and projections for 2006--2100. 

Our analysis primarily relies on ALADIN-5.3 (hereafter named ALADIN), a RCM developed by Météo-France~\cite{RN2738,RN2739}. To assess the robustness of our findings against the choice of this particular RCM, we replicate the analysis with three other RCM outputs, namely HIRHAM~5 (HIRAM, DMI~\cite{RN2740}), RACMO-22E (RACMO, KNMI~\cite{RN2742}), and REMO2009 (REMO, MPI~\cite{RN2741}). Unless otherwise specified, the results mentioned below are from ALADIN.

\subsection{Climate variables}
We base the analysis on the screen-level air temperature and consider its daily mean (T2m), minimum (T2m,min), and maximum (T2m,max). 

We first compare the evolution of the annual median of T2m (hereafter noted $\aoverbrace[L1R]{T2m}$) to that of the cold and hot quantiles (percentiles 1, 2, 5, 10, and 20, and percentiles 80, 90, 95, 98, and 99, respectively) of the annual PDF of T2m.
Consistent with the approach of the new normals~\cite{RN2732}, we consider the PDF as it evolves as a function of time rather than the historical PDF.

In order to assess whether our result depend on the definition of the central trend and the extremes, we also compare the evolution of the yearly averaged T2m (hereafter noted $\overline{T2m}$) with that of the values at $\pm0.5\sigma$, $\pm1\sigma$, $\pm1.5\sigma$, and $\pm2\sigma$, $\sigma$ being the yearly standard deviation of the daily T2m values. If T2m were to follow a Gaussian distribution, these thresholds would correspond to percentiles 30 and 70, 16 and 84, 6 and 94, and 2 and 98, respectively. 
We also consider the number of tropical nights (T2m,min $\ge \SI{20}{\celsius}$) and the number of frost days (T2m,max $\le \SI{0}{\celsius}$)~\cite{RN2794}. 

Finally, as the duration of extreme temperature events is a key aspect of their potential impacts on natural and human systems, we defined a warm days clustering index (WDCI), in an approach consistent with the percentile-based characterization of extremes used in this study. More specifically, we define the WDCI as the total number of days that belong to a sequence of at least three consecutive days with a daily-averaged T2m above the percentile 90 of the local annual T2m PDF. As the WDCI is calculated relative to the annual cycle, it mainly focused on the warm summer episodes. In order to account for the temporal evolution of the PDF with respect to global warming, we detrend the PDF by substracting a 30-year linear regression fit.
Tropical nights, frost days, and clustered warm days  occur in a limited number each year (e.g, WDCI $\le \SI{35}{yr^{-1}}$ according to the above definition). To achieve statistical significance, we bin their yearly counts into 30-years periods (1951--1980, 1981--2010, 2011--2040, 2041--2070, and 2071-2100).

\subsection{Evaluation of the shifting velocity}

An index of the velocity of temperature change has first been proposed by Loarie \emph{et al}.~\cite{Loarie2009}. Here, we follow the same approach, calculating the shifting velocity of a given value $\psi$ at each grid point as the ratio of its temporal derivative (denoted as temporal gradient by Loarie \emph{et al.}~\cite{Loarie2009}) to its gradient (denoted as spatial gradient by Loarie \emph{et al.}). Although intuitive, this definition relies on several implicit assumptions. We detail further this approach, its mathematical foundations, and we explicitly describe and discuss the underlying assumptions in the Supplementary Material.

At each grid point, the temporal average of the temporal derivative $\overline{\left(\frac{\partial \psi}{\partial t}\right)}$ is  determined as the slope of a linear regression on the time series over 1951--2100. The gradient is calculated with centered differences, in a way similar as Loarie \emph{et al}.~\cite{Loarie2009}. 
While the pace of climate change and forcings do not evolve linearly over the 150-years-long time span, the consideration  of a linear trend over this period provides a first-order assessment with minimal influence of the short-term fluctuations. Furthermore, a longer time period implies wider changes in both the central trends and temperature extremes, maximizing their chances to exhibit different behaviors.

\section{Results}
\subsection{Median and extreme quantiles}

\begin{figure}
\includegraphics[width=1\columnwidth, keepaspectratio]{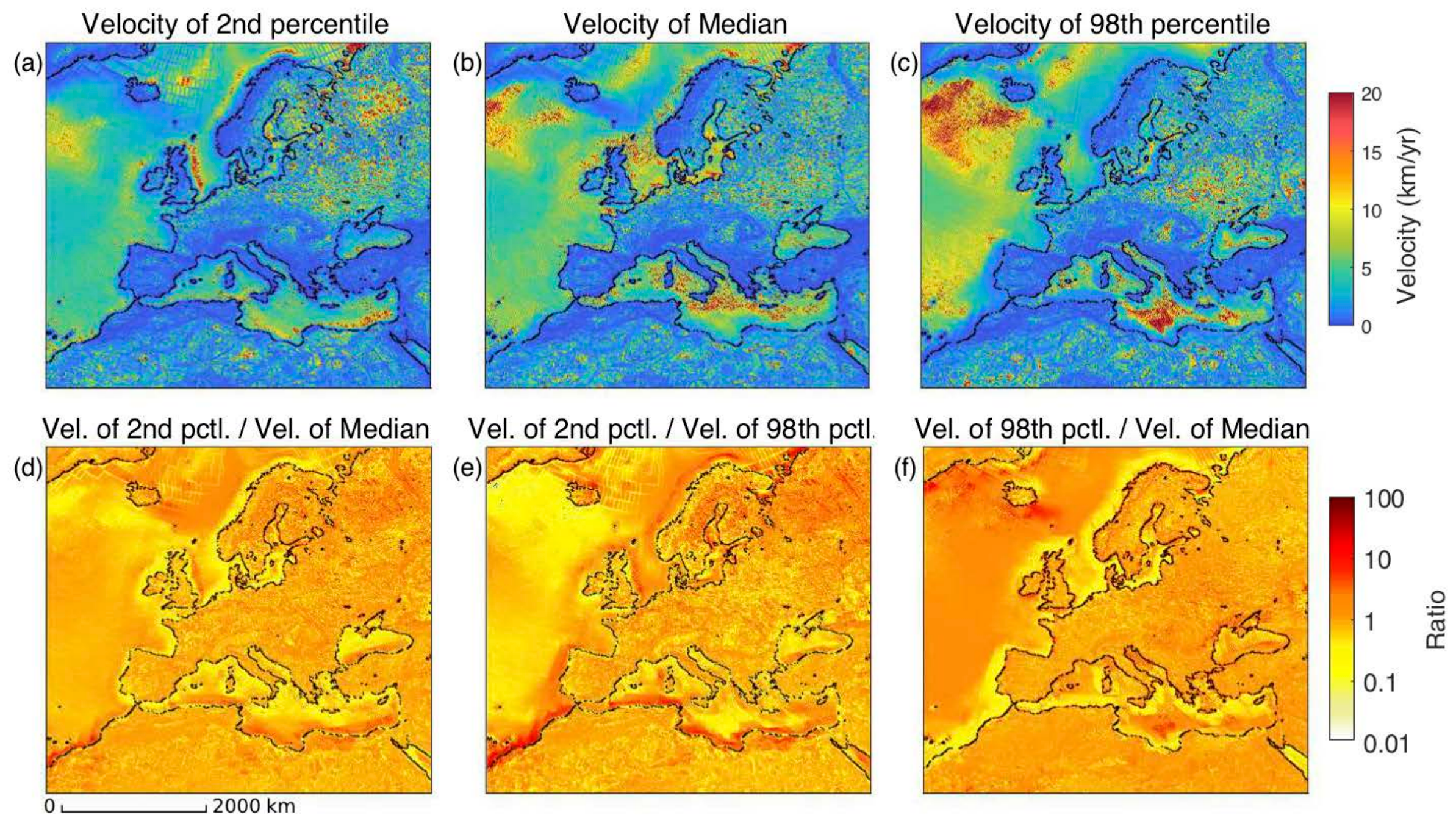}
\caption{(a--c) Shifting velocities averaged over 150 years (1951--2100) of percentiles 2 (low extremes, a) 50(median, b) and 98 (hot extremes, c) of the daily T2m. (d--f) Ratio of these velocities: (d) percentile 2 to median; (e) percentile 2 to percentile 98; (f) percentile 98 to median.}
\label{fig:median_quantiles}
\end{figure}

 \begin{figure}[h]
\includegraphics[width=\columnwidth, keepaspectratio]{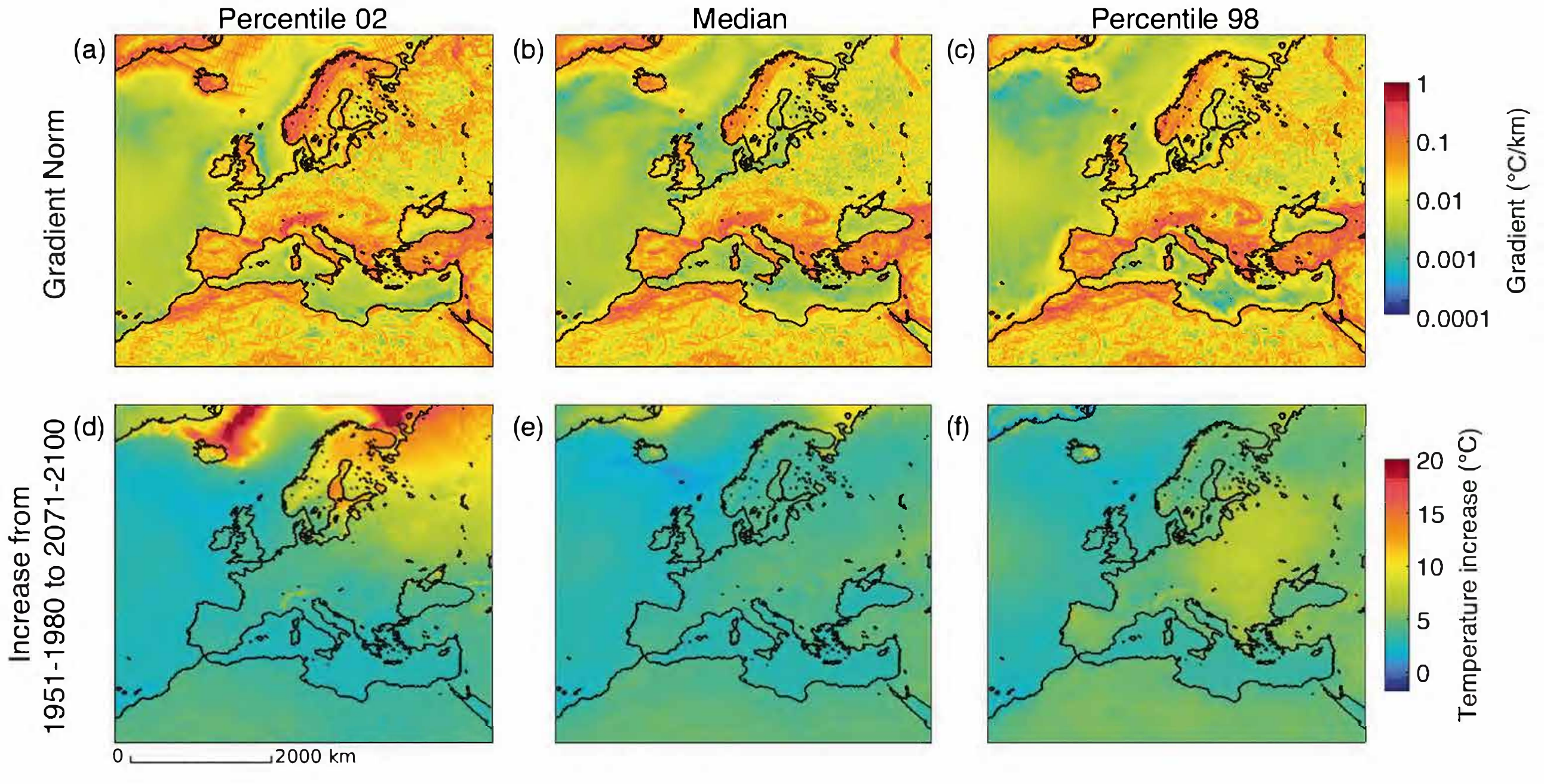}
\caption{(a--c) Gradients of T2m for percentiles 2 (a), 50 (b) and 98 (c), averaged over the period 1951--2100. (d--f) Evolution of percentiles 2 (d), 50 (e), and 98 (f) of the T2m PDF between the periods 1951--1980 and 2071--2100.}
\label{fig:gradient_percentiles}
\end{figure}

$\aoverbrace[L1R]{T2m}$ shifts faster over most marine areas, as expected from previous works~\cite{RN2751} (\Cref{fig:median_quantiles}b). This is especially true over the Atlantic Ocean and the Mediterranean Sea. A faster shift is also found across Eastern Europe and Western Russia.
Conversely, $\aoverbrace[L1R]{T2m}$ shifts slower over mountainous regions such as the Alps, the Atlas, as well as in Scandinavia.
These fast (slow) shifting velocities correspond to smoother (steeper) gradients (\Cref{fig:gradient_percentiles}b), while the temperature change is quite homogeneous over the considered time period (\Cref{fig:gradient_percentiles}e).

The shifting velocity of hot and cold extremes, represented by the percentiles 2 and 98 respectively (\Cref{fig:median_quantiles}a,c), show spatial patterns and magnitudes similar to those of the median, but differ at specific locations. The cold (hot) extremes shift slower (faster) than the median over the Atlantic Ocean West of Iceland and the Eastern part of the Mediterranean Sea. In contrast, they shift faster (slower) than the median on the East coast of Great Britain and Northern Russia. The similar shifting velocities of the median and the extremes appears to stem from a combination of both similar gradient (\Cref{fig:gradient_percentiles}a--c) and temporal derivatives (\Cref{fig:gradient_percentiles}d--f) on most of the considered region. However, the above-mentioned discrepancies seem to arise either from different gradients (West of Iceland, Eastern Mediterranean Sea, North Sea) or \gr{from} different temporal derivatives (Northern Russia). Finally, North of Iceland and in Eastern Europe, the  dependency of the gradients and temporal derivatives with regard to the quantiles compensate each other, resulting in similar shifting velocities.

The ratio between the shifting velocities of percentiles 2, 50, and 98 (\Cref{fig:median_quantiles}d--f) evidence further details pertaining to coastal regions. Off the coasts of Western Europe and in the North-Eastern Mediterranean Sea, both cold and hot extremes shift significantly slower than the median. In contrast, off the North coast of Africa, both in the Atlantic Ocean and the Southern Mediterranean Sea, the cold extremes shift faster than the median, while the hot extremes shift slower.

\begin{figure}
    \centering
    \includegraphics[width=\columnwidth, keepaspectratio]{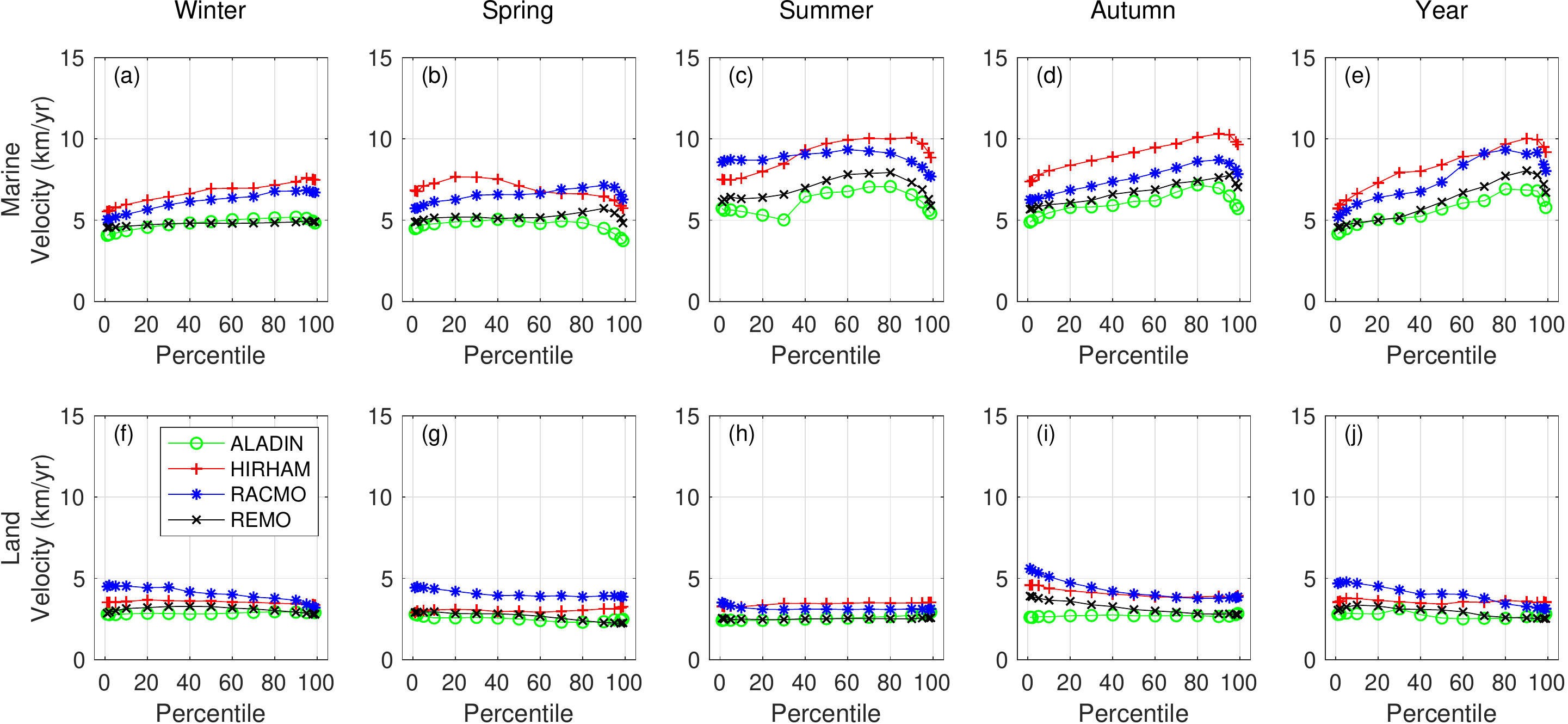}
    \caption{Mean shifting velocities as a function of the quantiles, averaged over marine regions (a--e) and land (f--j), for Winter (DJF, a and f), Spring (MAM, b and g), Summer (JJA, c and i), Autumn (SON, d and j), and for the yearly mean (e and k). Computations are done by four RCMs between 1951 and 2100, as detailed in the text and displayed in Panel f.}
    \label{fig:velocity_quantile_saisons}
\end{figure}

\begin{figure}
    \centering
    \includegraphics[width=\columnwidth, keepaspectratio]{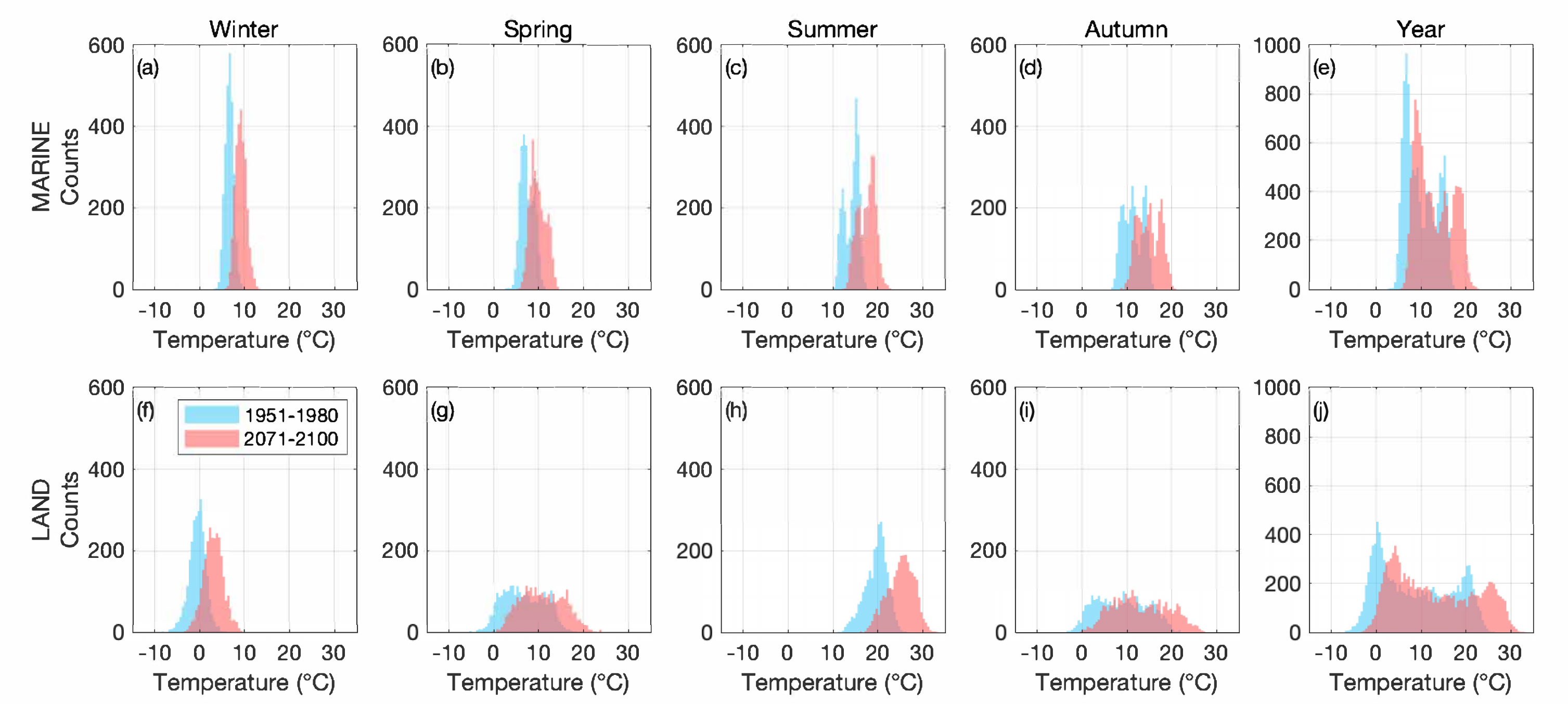}
    \caption{Evolution of the probability distribution functions of the daily mean T2m between 1951--1980 and 2071--2100, over marine regions (a--e) and over land (f--j), in Winter (DJF, a and f), Spring (MAM, b and g), Summer (JJA, c and i), Autumn (SON, d and j), and for the whole year (e and k); the vertical scale of the latter has been adapted for clarity.}
    \label{fig:PDF}
\end{figure}

\Cref{fig:velocity_quantile_saisons} displays the shifting velocity as a function of the quantile. The rather homogeneous velocities described above translate into very flat dependencies. As all quantiles evolve at the same pace, the histograms of the T2m PDF shift as a whole towards warmer temperature, for each season and on both marine regions and land as shown in \Cref{fig:PDF}. A faster (slower) heating of the hot (cold) extremes as compared with the median would have resulted in longer and fatter PDF tails at the end of the evolution period, while the opposite would have yielded sharper cutoffs of the tails.

Three alternative climate models (HIRHAM, RACMO, and REMO) \jk{generally} yield  similar trends as ALADIN in spite of local quantitative differences in the shifting velocities (\Cref{fig:modeles}). The shifting velocity increases with quantiles in all models in the Mediterranean Sea and the Atlantic Ocean especially West of Iceland. Also, the median shifts faster than both extremes in the North Sea and the Sea of Norway in the four models. In contrast, the decrease of the shifting velocity with the quantile over Russia is much stronger in RACMO than in ALADIN, and smaller in REMO and HIRHAM. \jk{As a result, the cold quantiles shift faster over land in RACMO than in the other models (\Cref{fig:velocity_quantile_saisons}j). This could arise from how Arctic warming influences cold outbreaks across Europe.} 
In all models, the differences in shifting velocity are mainly driven by the gradient patterns. Furthermore, while the overall shifting velocity may differ from model to model, the dependencies with the quantiles (i.e., the slope of the curves in \Cref{fig:velocity_quantile_saisons}) are \jk{mostly} similar.
While multi-model ensemble simulations would increase the accuracy and the delineation of uncertainty ranges~\cite{RN2807}, the similarity of the results across models indicates that relying on a single model is sufficient for the purpose of this study.

\begin{figure}[t]
\includegraphics[width=\columnwidth, keepaspectratio]{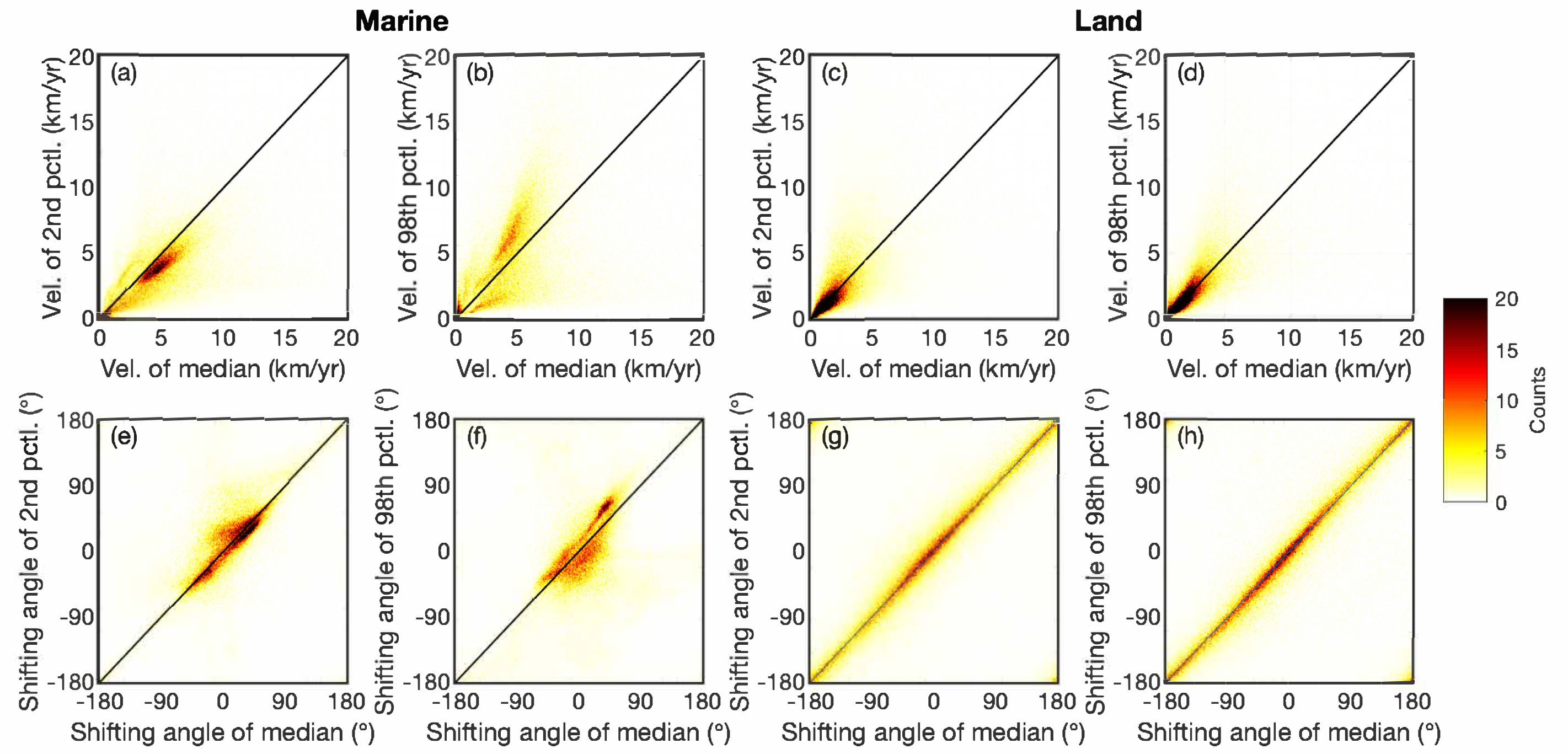}
\caption{Two-dimensional histograms of the distributions of the shifting velocities of T2m percentiles in Europe, over marine regions (a,b,e,f) and over land (c,d,g,h), between 1951 and 2100. Velocity magnitude (a--d) and direction (e--h) of percentiles 2 (a,c,e,g) and 98 (b,d,f,h), with respect to the median. The color scale indicates the amount of grid cells that display a pair of given velocities for the considered percentile. Directions \SI{0}{\degree}, \SI{90}{\degree}, \SI{\pm180}{\degree}, \SI{-90}{\degree} respectively refer to northward, eastward, southward and westward.}
\label{fig:median_correlations}
\end{figure}

\Cref{fig:median_correlations} displays two-dimensional histograms where the color scale indicates the amount of grid cells that display a pair of given velocities for $\aoverbrace[L1R]{T2m}$ and for the 2$^{nd}$ or 98$^{th}$ percentiles. As the shifting velocity $\vec{v}$ is a vector, Panels a--d focus on the velocity magnitude, while panels e--h display the shifting directions.
The diagonal corresponds to a perfect equality between the two shifting velocities: The clustering of data along this line illustrates the similar shifting velocities of the median and the extreme percentiles, as discussed above (\Cref{fig:median_quantiles}). This is particularly the case over land for both both magnitudes (\Cref{fig:median_correlations}c,d) and directions  (\Cref{fig:median_correlations}g,h).

Over marine regions, shifting velocities are slightly more dispersed. The percentiles 2 (\Cref{fig:median_correlations}a) and 98 (\Cref{fig:median_correlations}b) respectively display marginally slower and faster shifts than the median, consistent with the results showed in \Cref{fig:median_quantiles}. For cold extremes (Panel a), the stripe deviating down from the diagonal corresponds to coastal regions and to the Atlantic Ocean West of Iceland. Regarding hot extremes (Panel b), the stripe above the diagonal corresponds to the Atlantic Ocean.
In spite of these small fluctuations, the overall shifting directions of the extreme percentiles and the median over marine regions are rather similar, although slightly more spread than those over land. 

The same analysis has been carried out by considering the percentiles 1 (resp. 99), 5 (95), and 10 (90) instead of 2 (98) as the cold (hot) extremes. Similar results are found, although the features described above are slightly stronger for more extreme percentiles. 
We also performed the same analysis with the annual mean as the central value, and $\pm 0.5\sigma$, $\pm \sigma$, $\pm 1.5\sigma$, or $\pm 2\sigma$ of the T2m, as the temperature extremes  (\Cref{fig:mean_SD,fig:mean_direction,fig:gradients_SD} for $\pm 2\sigma$). The results are very similar with each other and with those obtained considering quantiles and the median.
This can be explained by considering that the yearly median and the mean of the daily T2m are extremely correlated, both in space ($R \ge 0.99$) and in time ($R \ge 0.999$). Their difference is limited to \SI{\pm0.5}{\celsius} over almost 70\% of the grid cells. 

Therefore, the behavior of the shifting velocity of the temperature extremes defined with respect to the T2m PDF is to a large extent resilient to the particular definition of the extremes. This immunity covers the choice of the central trend  (median or mean), the tails of the distribution (percentiles or standard deviations), as well as the cutoff chosen in the temperature distribution to define the extremes. This corroborates the robustness of our findings.

\subsection{Tropical nights, frost days, and warm days clustering}

\begin{figure}
\includegraphics[width=\columnwidth, keepaspectratio]{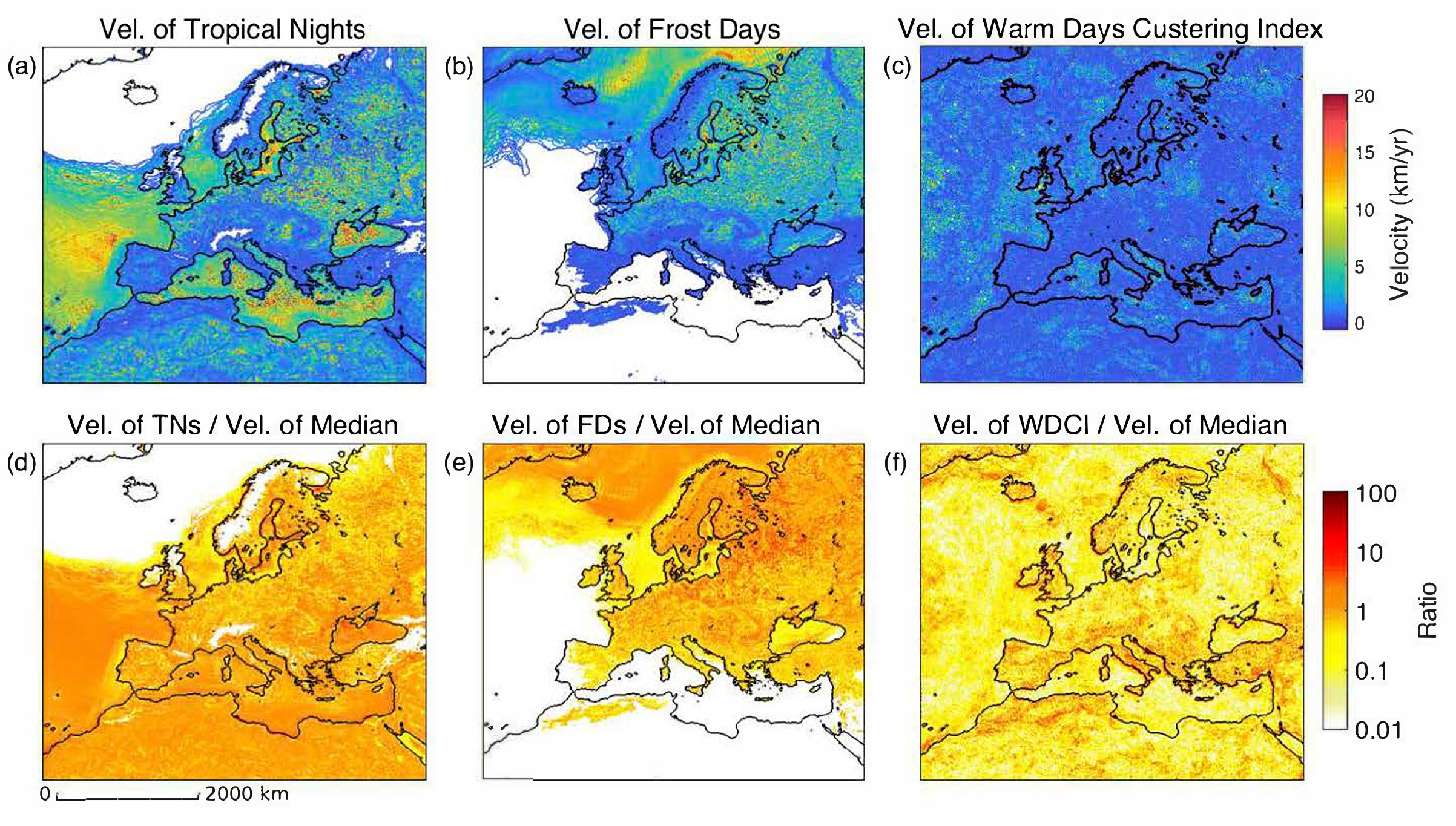}
\caption{(a,b,c) Shifting velocities of the occurrence frequency of (a) tropical nights (TNs), (b) frost days (FDs), and (c)  Warm days clustering index (WDCI). (d--f) Corresponding ratio to the shifting velocities of the median T2m. Sub-regions where the indices do not apply are displayed in white.}
\label{fig:heat_waves}
\end{figure}

\begin{figure}
\includegraphics[width=\columnwidth, keepaspectratio]{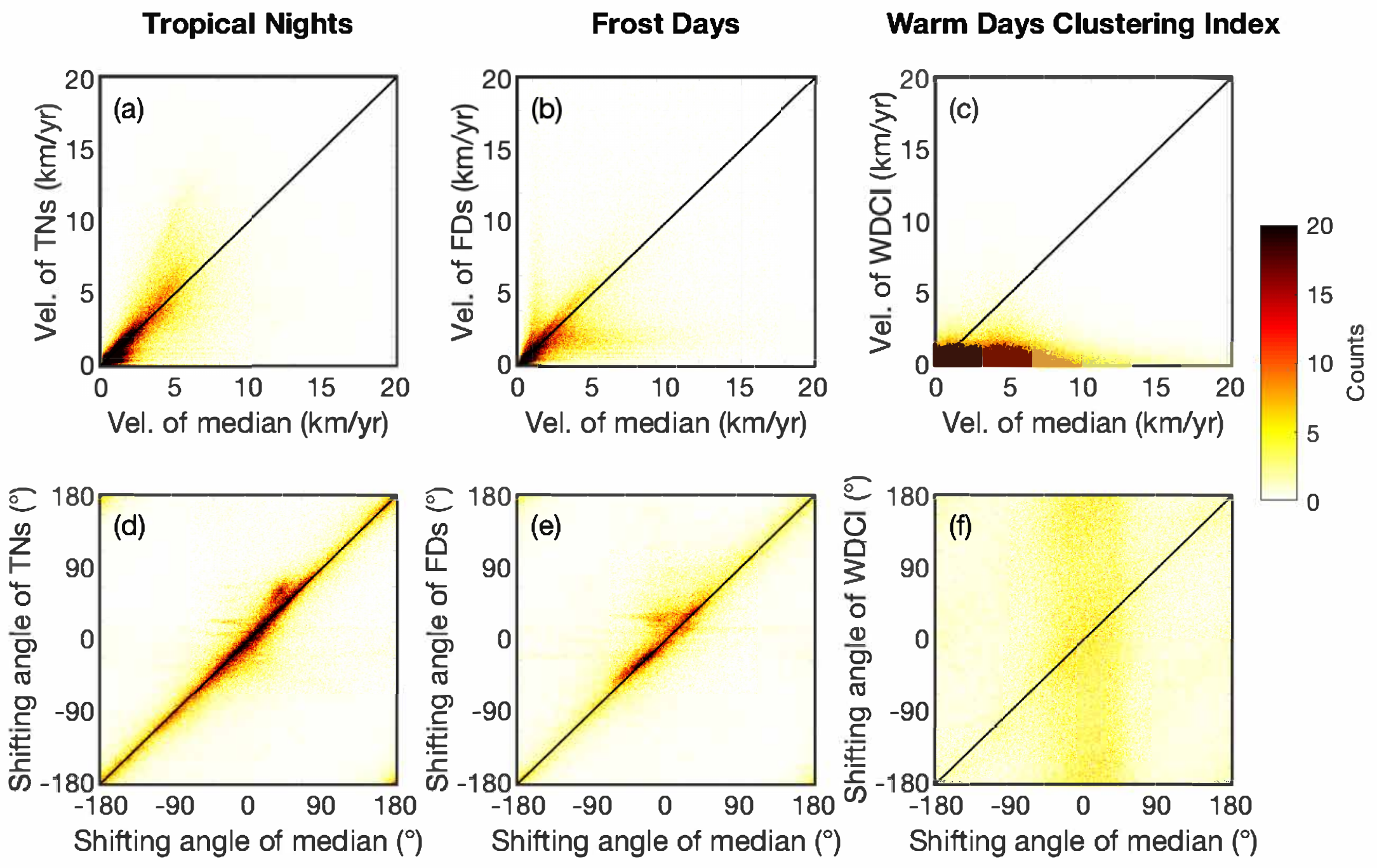}
\caption{Two-dimensional histograms of (a--c) the magnitude and (d--f) direction of the shifting velocities of (a,d) tropical nights, (b,e) frost days, and (c,f) WDCI compared to the median T2m, over the period 1951--2100. Directions \SI{0}{\degree}, \SI{90}{\degree}, \SI{\pm180}{\degree}, \SI{-90}{\degree} respectively refer to northward, eastward, southward and westward.}
\label{fig:extremes_direction}
\end{figure}

Isochrones defining the number of tropical nights per year generally shift much faster over marine regions, as well as over Russia and Northern Africa (\Cref{fig:heat_waves}a).
The behavior is similar to that of $\aoverbrace[L1R]{T2m}$ (\Cref{fig:median_quantiles}b), in spite of a slightly faster shifting velocity. As a result, the ratio between the corresponding velocities is quite homogeneous, with an average value of 1.08 over all regions where it is defined (\Cref{fig:heat_waves}d). The slightly faster shifting velocity of the tropical nights as compared with $\aoverbrace[L1R]{T2m}$ is also evidenced by a deviation of data above the diagonal in the two-dimensional histograms of \Cref{fig:extremes_direction}a.
The only exceptions are the Atlantic Ocean off the Spanish coast and the Black Sea, where the tropical night isochrones locally shift up to 2--6 times faster than the median. 
In spite of these deviations on the velocity norm, the number of tropical nights shifts in the same direction as $\aoverbrace[L1R]{T2m}$ (\Cref{fig:extremes_direction}d). The dispersion of shifting directions mostly corresponds to the rounding errors related to low shifting velocities. 

Where they occur, the number of frost days also shift at velocity magnitudes and directions comparable to that of $\aoverbrace[L1R]{T2m}$ (\Cref{fig:heat_waves}b, \Cref{fig:extremes_direction}b,e) over most of the domain. However, they shift slower than the median in both the North Sea and in the Atlantic Ocean south of Iceland (\Cref{fig:heat_waves}e). 

 Finally, the WDCI shifts typically 4 times slower (\SI{1}{km/year}) over the whole domain than $\aoverbrace[L1R]{T2m}$ as well as than the other extremes investigated in this study (\Cref{fig:heat_waves}c,f and \Cref{fig:extremes_direction}c).
 Slightly faster shifts are however observed over the Atlantic Ocean (\SI{1.7}{km/year} on average) and to a lesser extent over Russia (\SI{1.2}{km/year} on average).

\section{Discussion}

The results reported above show no systematic differences, neither in direction nor in magnitude, between shifting velocity vectors of the central temperature trend, and that of the extremes. This is especially true over land and with regard to the extremes defined relative to the temperature PDF. This can be understood by considering that the temperature PDF turns out to be to a large extent spatially homogeneous, and that its shape barely evolves as it shifts to higher temperatures as climate warms up.

This study also shows that the shifting velocity conveys complementary information as compared to the local temperature evolution. In Scandinavia, Northern Russia, and offshore from them, the cold extremes increase much faster than hot extremes (\Cref{fig:gradient_percentiles}d,f). However, as their gradients are also steeper (\Cref{fig:gradient_percentiles}a,c), this does not translate into faster shifting velocities. Steep gradients can be related to the edge of the sea-ice or snow cover. As long as they are present, ice and snow packs keep cold extreme values below the freezing point of water. Conversely, ocean and land surfaces are much less reflective with respect to solar radiation, ensuing positive ice-snow albedo feedbacks~\cite{RN2793}. As a consequence, displacements of the edge of the ice or snow cover locally allows the cold extreme to abruptly rise beyond $ \SI{0}{\celsius}$.

The finding that temperature extremes and the central trend generally shift at a similar velocity could appear to contradict previous studies, which found extremes to actually increase faster than the central trends~\cite{RN2735,RN2736,RN2734}. This apparent paradox is related to the definition of the extremes. 
Here we consider extremes as the tails of the temperature PDF \emph{during separate time periods}, while most studies define extremes with regards to the historical ($e.g.$, the 1990s) PDF. As the PDF shifts as a whole, an increasing number of climatic conditions falls into the definition of extremes relative to today's or past PDF. They however become new normals~\cite{RN2732} since their values approach the new central ones. 
In other words, today's hot extremes would become increasingly frequent, but also increasingly "normal", $i.e.$, less and less extreme from a statistical point of view. In that sense, today's extremes draw an idea of tomorrow's central trends, but under such a definition, extremes will not become more frequent nor further from the normal in the future.

Temperature extremes defined with reference to absolute thresholds also generally shift as a whole at a comparable velocity as the central trends, again with local exceptions. Indeed, one may note that T2m,min and T2m,max shift at a same velocity as the T2m itself (\Cref{fig:shift_T2mMinMax}; See also \Cref{fig:median_quantiles}b), \emph{i.e.}, the daily thermal amplitude does not increase with climate warming over the considered region. Consequently, the threshold conditions also shifts together with the central trend.

Only the WDCI displays a shifting velocity significantly different (and much slower) from that of $\aoverbrace[L1R]{T2m}$ and the other indicators investigated in this work. The reason for this is that WDCI represents a metric of the clustering of the number of days with T2m in the hottest decile. The yearly number of days in the hottest decile is constant by definition. The very slow average shifting velocity (1 km/yr, except over the Atlantic Ocean) of the WDCI isochrones simply means that they do not tend to cluster into longer exceptionally hot periods.

\section{Conclusion}
In this paper, we calculate the shifting velocities of central trends and extremes of screen-level air temperature, using EURO-CORDEX RCM simulations between 1951 and 2100, according to the RCP8.5 emission scenario. \jk{We mostly focus on the ALADIN model in order to support a detailed analysis.} We find that T2m extremes shift at a comparable velocity as the central trends over Europe, except over a limited number of sub-regions. This somewhat unexpected result can be explained on the basis of the T2m PDFs that generally shift as a whole with little deformations.  Consequently, today's extreme situations would, to a large extent, become new normals by the end of the $21^\textrm{st}$ Century. While current hot extremes will become more and more frequent, these situations will not be deemed extreme anymore in a warmer climate.  Similarly, cold extremes will not disappear under climate warming, but they will be less and less cold as the normals warm up.

Research shows that biodiversity adaptability to increased temperatures is somewhat limited and that their first response to changing climatic conditions is a shift in location~\cite{RN2801}, with migration patterns often being slower than changes in climatic conditions~\cite{RN2804}. Overall, human adaptability to temperature extremes \sg{has improved} over the past decades~\cite{RN2802,RN2805}, but further research is needed to compare the speed of human adaptability with that of climate shift and to characterize context-specific limits of human adaptability~\cite{RN2803,RN2800}. A faster shift of extremes as compared to the normals would have implied specific adaptation issues. Conversely, our finding that extremes shift at a similar velocity as the normals may limit their impact on the adaptability of ecological and human systems.

Further assessments regarding shifting velocities should be carried out by \jk{performing more detailed inter-model comparisons}, taking into account the deformation of isotherms during their displacements, and investigating the role of processes like soil moisture, snow and sea-ice modifications and related surface-atmosphere feedbacks. 
\jk{The consideration of climate variables other than the screen-level air temperature, as well as other regions of the world}, will also allow help to assess the generality of our conclusions. In particular, a focus on marginal ice regions at high latitudes and areas where climate change substantially affects land use would allow to investigate the effects of multi-stability and  threshold effects on the climate shifting velocity.
Further research could also develop ways to assess the velocity of composite \jk{indices} (e.g. made of temperature and precipitation variables) in order to explore the simultaneous shift of different climate extremes.

\textbf{Acknowledgments}. 
We gratefully acknowledge technical support by Roman Kanala at the Institute for Environmental Sciences. The EURO-CORDEX data used in this study were obtained from the Earth System Grid Federation server (https://esgf-data.dkrz.de/projects/esgf-dkrz/). We thank all the modeling groups that performed the simulations and made their data available. \gr{NCAR is supported by the National Science Foundation}

\bibliographystyle{iopart-num}
\bibliography{guillaumebiblio,biblio_velocity}

\newpage
\section*{Supplementary Material}

\setcounter{figure}{0}

\renewcommand{\thefigure}{S\arabic{figure}}

\subsection*{Supplementary Methods}

\begin{figure}[h!]
    \centering
    \includegraphics[width=0.6\columnwidth,keepaspectratio]{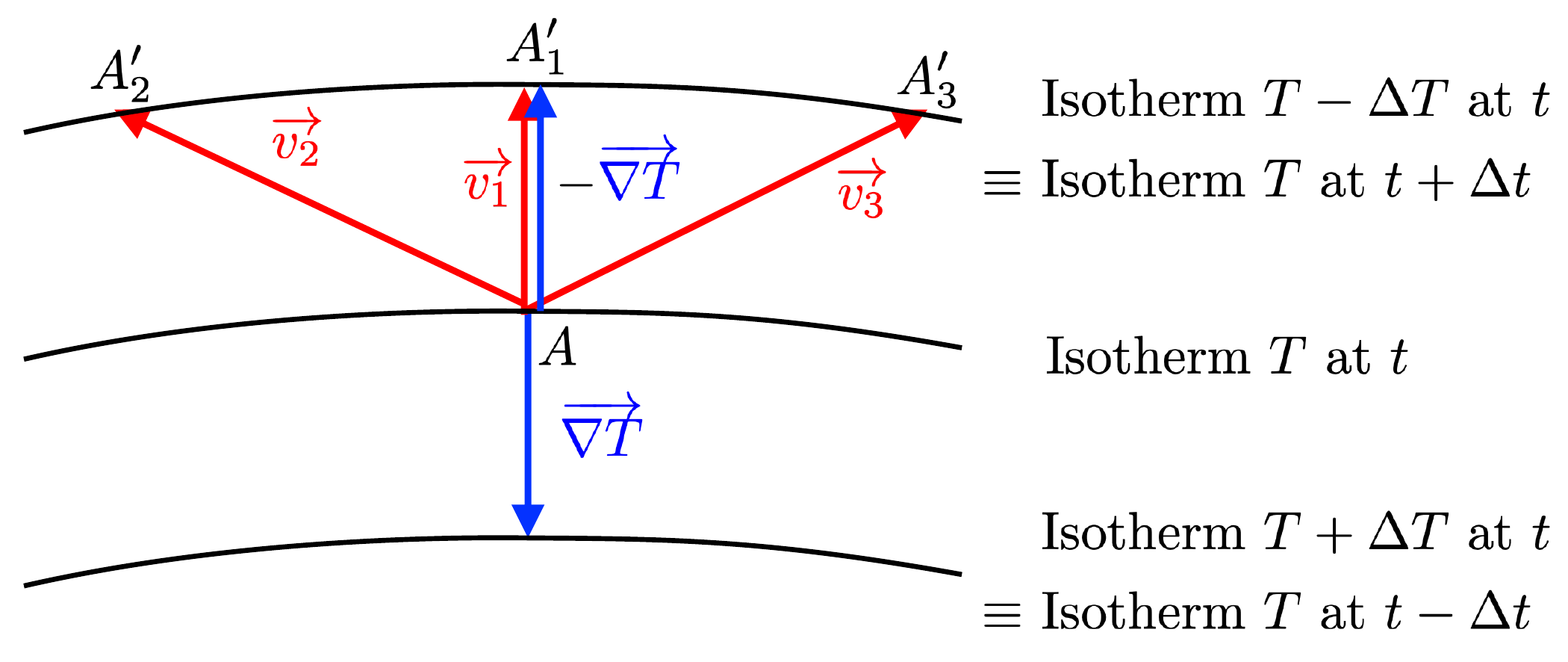}
    \caption{Ambiguity of the velocity determination. The climate change, here illustrated as global warming, translates into a northward-shifting of the isotherms. In this process, a point $A$ located on the isotherm $T$ at time $t$ will be displaced to a location $A'$ on the new isotherm $T$ of time $t+\Delta t$. However, its location on the new isotherm is unspecified. Here, three examples of possible arrival spots, $A'_1$, $A'_2$, and $A'_3$, are displayed. They respectively correspond to velocities $\vec{v_1}$, $\vec{v_2}$, and $\vec{v_3}$, respectively, that have different norms  and directions. Under the assumptions detailed in the text (in particular, negligible isotherm deformation). By definition, the gradient $\vec{\nabla T}$ (blue vector) is perpendicular to the isotherms. Therefore the displacement antiparallel to $\vec{\nabla T}$ (i.e., leading to $A'_1$, has the shortest magnitude, and corresponds to the slower shifting velocity $\vec{v_1}$. This is the most natural and therefore almost universal choice, that however needs to be explicited and justified, as discussed in the text of this Supplementary Methods.}
    \label{fig:indetermination}
\end{figure}

While the calculation of the shifting velocity of isopleths as the ratio of the temporal derivative to the gradient appears intuitive, it deserves justification, however. Furthermore, this ratio defines at most the magnitude of the velocity vector, but does not provide information on its direction, which has to be determined on the basis on assumptions that are usually kept implicit as they are implemented in GIS functions.

In this Section, we discuss these aspects in detail in order to provide the reader with a rigorous justification of the methods used in this work as well as previous ones, including Loarie et al.~\cite{Loarie2009}.

The starting point of the discussion is the well-known relation between the Eulerian (or partial) derivative $\frac{\partial \psi}{\partial t}$ and its Lagrangian (or particulate) counterpart $\frac{\textrm{d}\psi}{\textrm{d}t}$ of a value $\psi$ along the trajectory of an arbitrary point moving at a vectorial velocity $\vec{v}$~\cite{RN2756}:
\begin{equation}
\frac{\textrm{d}\psi}{\textrm{d}t} =\frac{\partial \psi}{\partial t} + \vec{v}\cdot\vec{\nabla}\psi .
\end{equation}
In the present work, $\psi$ may represent the yearly median or mean of T2m, the quantiles of the yearly T2m distribution, or the values at $\pm \lambda \sigma$, $\lambda$ being a positive real number. It may also be the numbers of tropical nights, of frost days, or the WDCI, binned over 30-years periods. 

An isopleth for a variable $\psi$ is defined as an iso-$\psi$ line. Therefore, the velocity of climate change is the shifting velocity of the relevant isopleth. As the isopleth has a constant value of $\psi$, the Lagrangian derivative is $\frac{\textrm{d}\psi}{\textrm{d}t} = 0$, for any point shifting on the isopleth. As a consequence: 

\begin{equation}
\frac{\partial \psi}{\partial t} = -\vec{v}\cdot\vec{\nabla}\psi .
\label{eq:Euler_Lagrange}
\end{equation}

As a vector, $\vec{v}$ has at least two components: say its coordinates along two horizontal axes, or, equivalently, its norm $||\vec{v}||$ and its direction, characterized by an angle $\theta$ with regard to the North. However, \Cref{eq:Euler_Lagrange} is scalar, so that a second relationship between $\vec{v}$ and $\vec{\nabla}\psi$ is required to determine $\vec{v}$. 
This indetermination of the velocity vector is illustrated in \Cref{fig:indetermination}. 
The most natural and common practice, although generally implicit, is to follow a the analogy of a mechanical motion on a slope, that occurs along the maximal slope line, i.e., in the direction opposite to the level sets, i.e., to the iso-altitude lines. Consequently, in this framework, it is assumed that the $\psi$-isopleths 
($e.g.$, isotherms if $\psi$ is temperature) shift down the gradient of $\psi$. Their velocity is therefore oriented opposite to the unit vector pointing along the gradient of $\psi$: $ \vec{u}_\nabla \equiv \frac{\vec{\nabla}\psi} {||\vec{\nabla}\psi||} $.
This assumption implies that (i) isopleths shift parallel to each other, (ii) their deformation is minimal, and (iii) their curvature can locally be neglected. Although quite stringent, this triple assumption is wide-spread as it is necessary to overcome the indetermination that stems from \Cref{eq:Euler_Lagrange}. The shifting velocity vector then expresses as:

\begin{equation}
    \vec{v}= - \frac{1}{||\vec{\nabla}\psi||} \frac{\partial \psi}{\partial t}\vec{u}_\nabla 
    \label{eq:v_vecteur}.
\end{equation}

The magnitude of the local shifting velocity $v = ||\vec{v}||$ of the isopleth, is then:

\begin{equation}
v = \frac{1} {||\vec{\nabla} \psi||} \frac{\partial \psi}{\partial t}.
\end{equation}

\Cref{eq:v_vecteur} averages over time to:
\begin{equation}
    \overline{\vec{v}}=-\overline{\frac{1}{||\vec{\nabla}\psi||} \frac{\partial \psi}{\partial t} \vec{u}_\nabla} \label{eq:v_vecteur_moyen}.
\end{equation}

$\vec{u}_\nabla$ is orthogonal to the isopleths, which are assumed to shift parallel to each other and without deformation. It can therefore be considered as constant, so that:

\begin{equation}
    \overline{\vec{v}}=-\overline{\frac{1}{||\vec{\nabla}\psi||} \frac{\partial \psi}{\partial t}} \vec{u}_\nabla
    = - \left[\frac{1}{\left|\left|\overline{\vec{\nabla}\psi}\right|\right|}\  \overline{\left(\frac{\partial \psi}{\partial t}\right)}
    +\mathrm{Cov}\left(\frac{1}{||\vec{\nabla}\psi||},\frac{\partial \psi}{\partial t}\right) \right] \vec{u}_\nabla
    \label{eq:v_vecteur_moyen2}
\end{equation}
where Cov denotes the covariance.
The assumption that the isopleths shift down the gradient, parallel to each other and without deformation implies that the covariance term is neglected, so that:

\begin{equation}
    \overline{\vec{v}}=-\frac{1}{\left|\left|\overline{\vec{\nabla}\psi}\right|\right|}\  \overline{\left(\frac{\partial \psi}{\partial t}\right)}\ \vec{u}_\nabla 
    \label{eq:v_vecteur_moyen2}.
\end{equation}

As a result, the magnitude of the average shifting velocity vector over the considered time period, hereafter denoted the \emph{shifting velocity}, expresses as:

\begin{equation}
\overline{v} = \frac{1}{\left|\left|\overline{\vec{\nabla} \psi}\right|\right|}\overline{\left(\frac{\partial \psi}{ \partial t}\right)}.
\end{equation}

\newpage
\subsection*{Supplementary figures}

\begin{figure}
    \centering
    \includegraphics[width=\columnwidth, keepaspectratio]{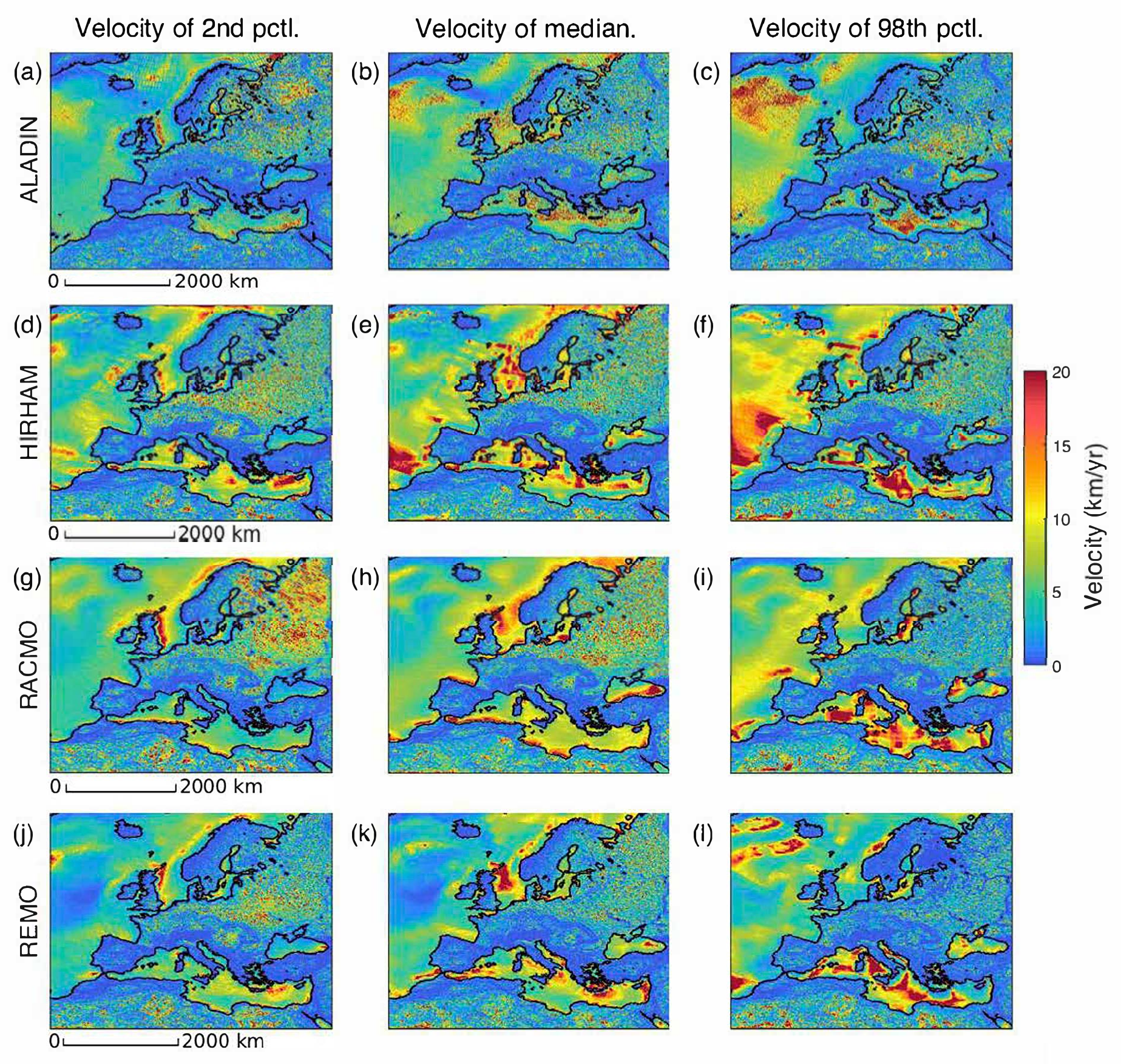}
    \caption{Shifting velocities over 150-year (1951--2100) of the daily T2m percentiles 2 (a,d,g,j), 50 (median, b,e,h,k) and 98 (c,f,j,l), for models ALADIN (from \Cref{fig:median_quantiles}, a--c), HIRAM  (d--f), REMO (g--i), and RACMO (j--l).}
    \label{fig:modeles}
\end{figure}

\begin{figure} [h]
\includegraphics[width=\columnwidth, keepaspectratio]{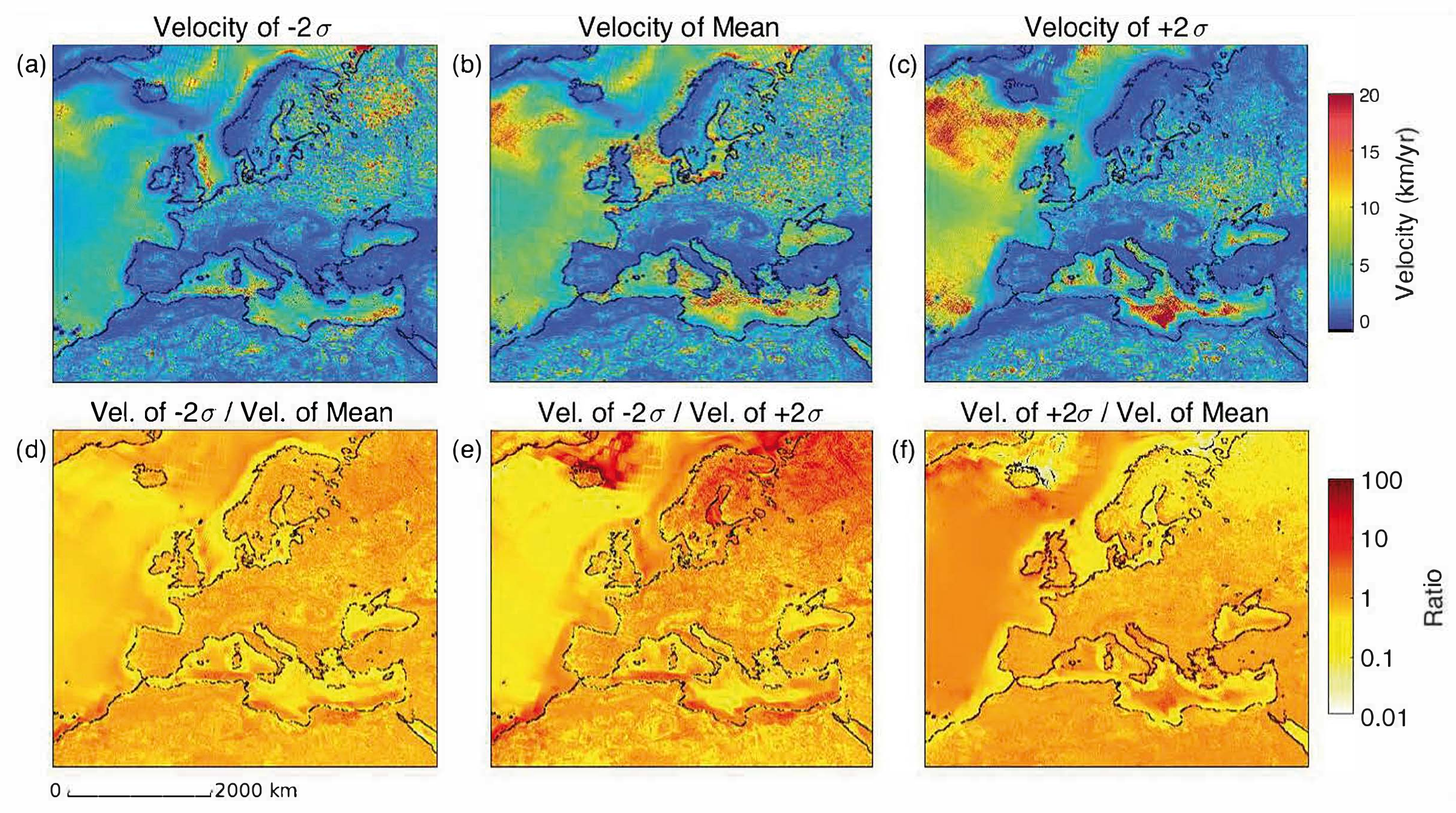}
\caption{(a--c) Shifting velocities of the daily T2m mean (b) and values at $-2\sigma$ (a) and  $+2\sigma$ (c). (d--f) Ratio of these velocities. (d) $-2\sigma$  to mean; (e) $-2\sigma$  to $+2\sigma$ (f) $+2\sigma$ to mean.}
\label{fig:mean_SD}
\end{figure}

\begin{figure}[h]
\includegraphics[width=\columnwidth, keepaspectratio]{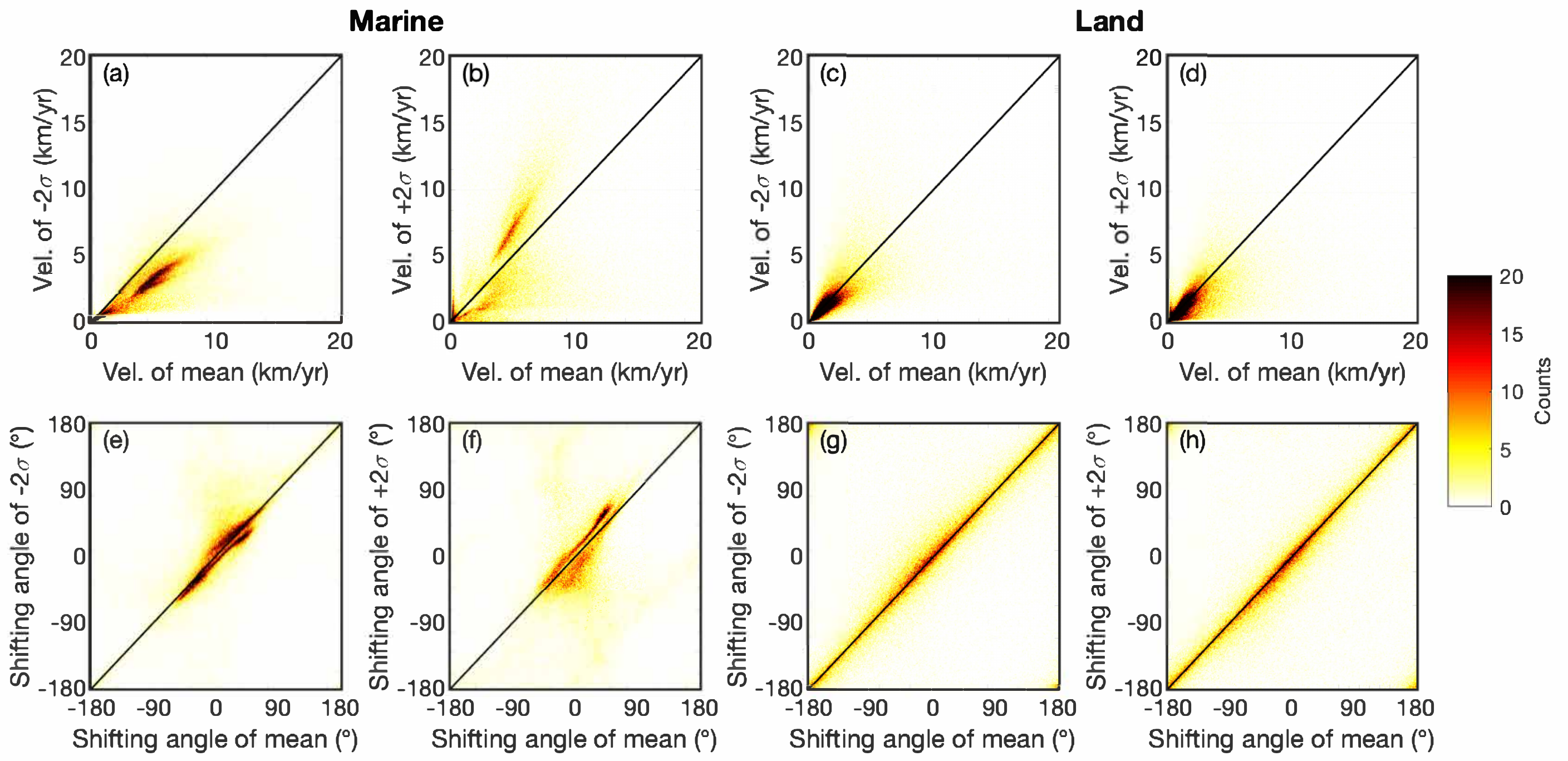}
\caption{Two-dimensional histograms of the distributions of the  shifting velocities of T2m mean and standard deviations in Europe, over marine regions (a,b,e,f) and over land (c,d,g,h), between 1951 and 2100. Velocity magnitude (a--d) and direction (e--h) of $-2\sigma$ (a,c,e,g) and $+2\sigma$ (b,d,f,h), with respect to the mean. Directions \SI{0}{\degree}, \SI{90}{\degree}, \SI{\pm180}{\degree}, \SI{-90}{\degree} respectively refer to northward, eastward, southward and westward.}
\label{fig:mean_direction}
\end{figure}

\begin{figure}[h]
\includegraphics[width=\columnwidth, keepaspectratio]{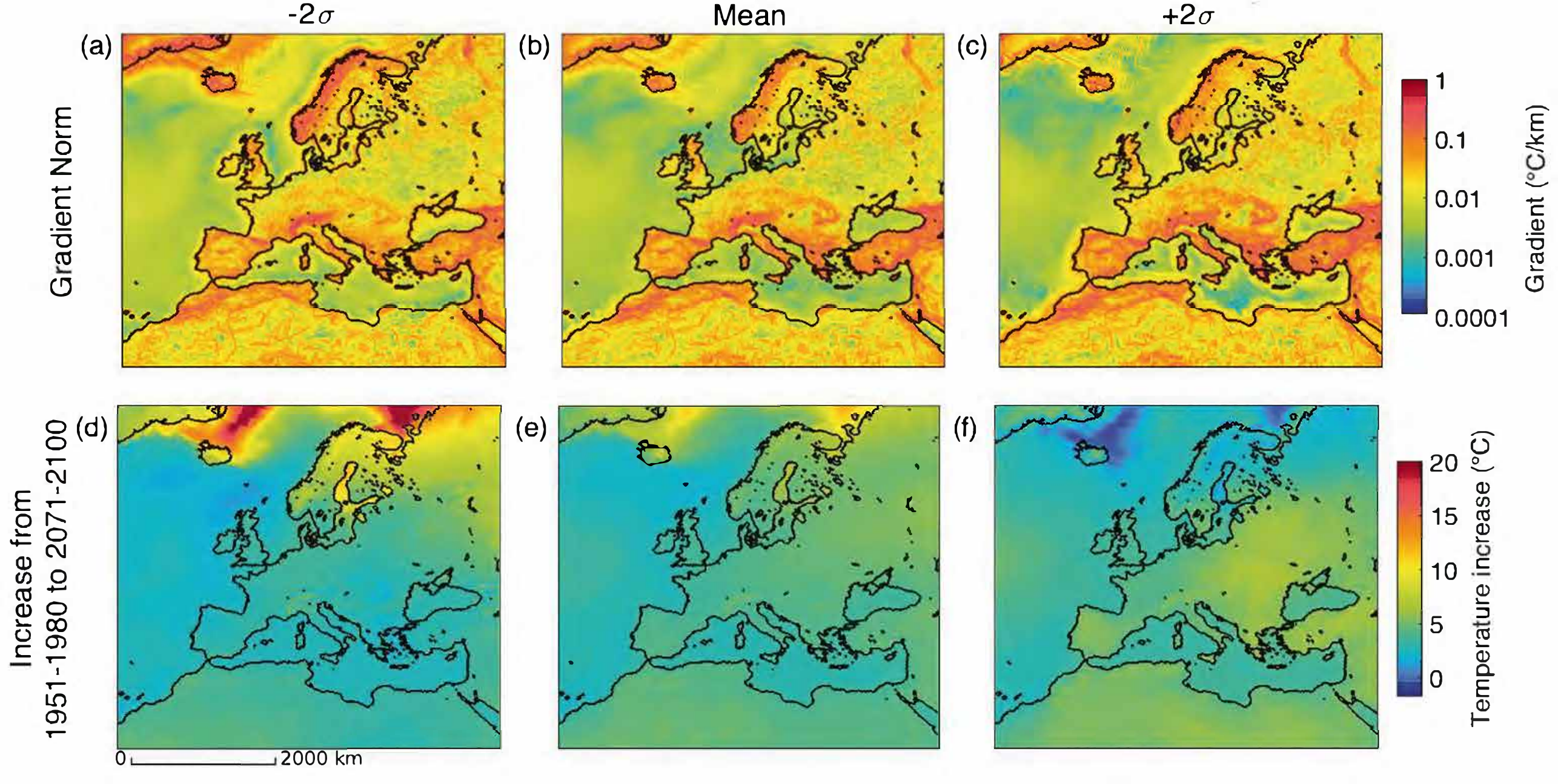}
\caption{(a--c) Gradients of T2m for $-2\sigma$ (a), mean (b) and $+2\sigma$ (c), averaged over the period 1951--2100. (d--f) Evolution of $-2\sigma$ (d), mean (e), and $+2\sigma$  (f) of the T2m PDF between the periods 1951--1980 and 2071--2100.}
\label{fig:gradients_SD}
\end{figure}

\begin{figure} [h]
    \centering
\includegraphics[width=0.7\columnwidth, keepaspectratio]{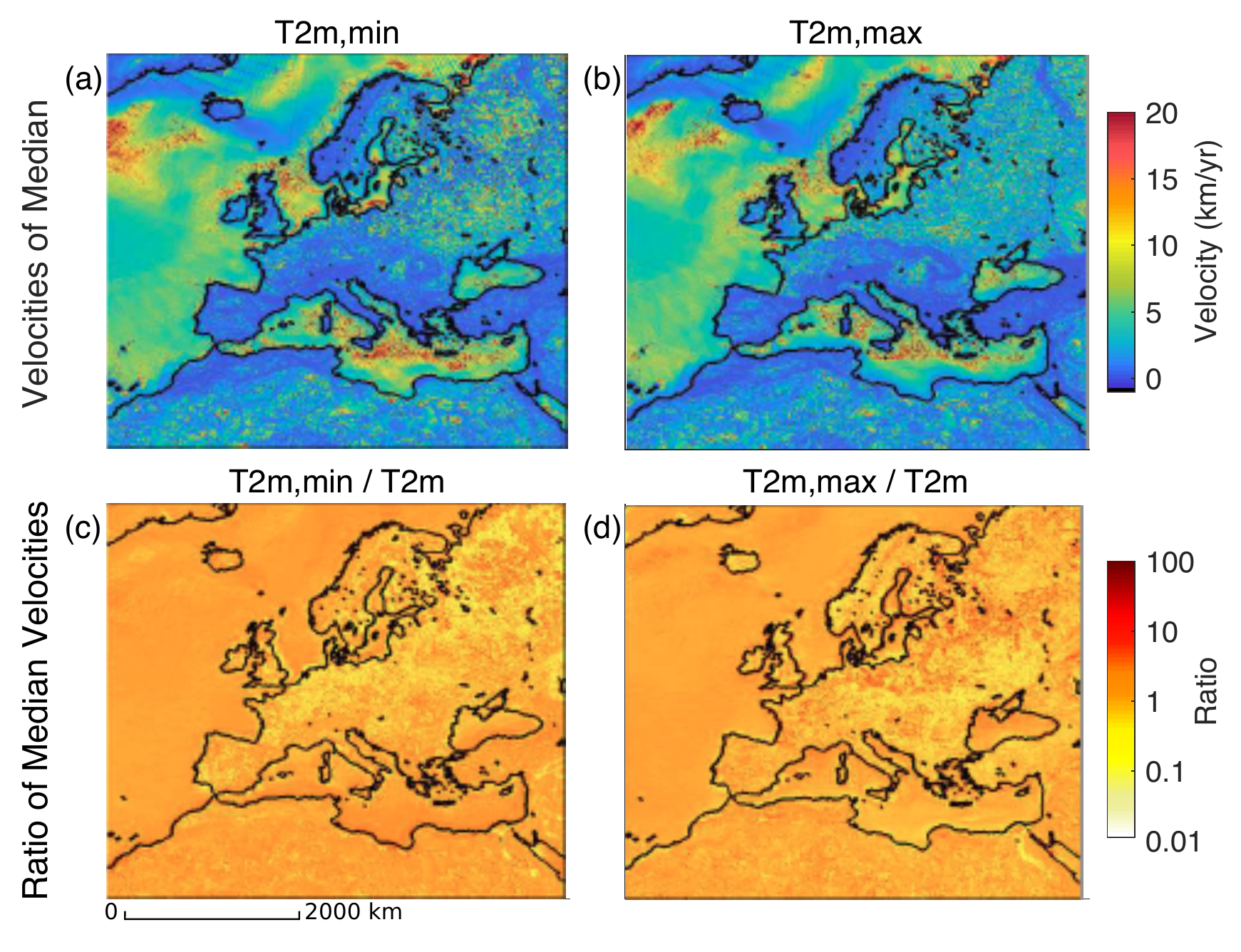}
\caption{(a,b) Shifting velocities of the daily T2m,min (a) and T2m,max (b). (c,d) Ratio of these velocities to that of the T2m median: (c) T2m,min, (d) T2m,max.}
\label{fig:shift_T2mMinMax}
\end{figure}

\end{document}